\documentclass[conference]{IEEEtran}
\IEEEoverridecommandlockouts

\usepackage[T1]{fontenc}
\usepackage[utf8]{inputenc}
\usepackage{cite}
\usepackage{amsmath,amssymb,amsfonts,bm,mathtools}
\usepackage{algorithmic}
\usepackage{array}
\usepackage{booktabs,tabularx,makecell}
\usepackage{adjustbox}
\usepackage{graphicx}
\usepackage{textcomp}
\usepackage{xcolor}
\usepackage{url}
\usepackage{flushend}

\allowdisplaybreaks

\title{Fluid Antenna Array-Inspired Location-Posterior-Driven Subarray Sizing and Power Control for Two-Hop AF UAV Relaying
}
\author{
\IEEEauthorblockN{Xuanyi Zhu\IEEEauthorrefmark{1}, Jian Dang\IEEEauthorrefmark{1}\IEEEauthorrefmark{2}, Chen Zhao\IEEEauthorrefmark{2}, Huaifeng Shi\IEEEauthorrefmark{2}, and Zaichen Zhang\IEEEauthorrefmark{1}\IEEEauthorrefmark{4}}
\IEEEauthorblockA{\IEEEauthorrefmark{1}National Mobile Communications Research Laboratory,\\
Frontiers Science Center for Mobile Information Communication and Security,\\
Southeast University, Nanjing 211189, China\\
E-mail: zhuxuanyi@seu.edu.cn; dangjian@seu.edu.cn; zczhang@seu.edu.cn}
\IEEEauthorblockA{\IEEEauthorrefmark{2}Key Laboratory of Intelligent Support Technology for Complex Environments,\\
Ministry of Education, Nanjing University of Information Science and Technology,\\
Nanjing 210044, China\\
E-mail: 002912@nuist.edu.cn; shihuaifeng@nuist.edu.cn}
\IEEEauthorblockA{\IEEEauthorrefmark{4}Purple Mountain Laboratories, Nanjing 211111, China}
\thanks{This work is supported by the Fundamental Research Funds for the Central Universities (2242022k60001), Basic Research Program of Jiangsu (No. BK20252003), the Key Laboratory of Intelligent Support Technology for Complex Environments, Ministry of Education, Nanjing University of Information Science and Technology (No. B2202402). (Corresponding Author: Jian Dang)}
}

\begin{document}
\maketitle

\begin{abstract}
	This paper develops fluid antenna array (FAA)-inspired subarray sizing and transmit-power design for a two-hop amplify-and-forward (AF) unmanned aerial vehicle (UAV) relay using progressively contracting user-location posteriors. A contiguous reconfigurable subarray is shared by first-hop reception and second-hop forwarding, such that its active size jointly determines the receive gain, forwarding gain, and beamwidth. By adaptively controlling the effective aperture, the proposed design exploits geometric reconfigurability to balance array gain against pointing robustness under location uncertainty. Projecting the position covariance onto the array direction yields a closed-form direction-limited size inversely proportional to directional uncertainty. Posterior samples are propagated through the two-hop rate model, and the subarray size and transmit power are then selected to minimize UAV power subject to a worst-user lower-tail rate requirement and hardware power limits. The planned configuration is further audited over instantaneous two-hop Rician channels at the true user positions. At $t=8~\mathrm{s}$, the proposed design saves $3.17$ dB over full-array narrow-beam transmission on paired feasible geometries and achieves $60.0\%$ service success at a $0.15$-W budget, compared with $43.2\%$ for a fixed eight-element subarray.
\end{abstract}


\section{Introduction}
\label{sec:intro}

Unmanned aerial vehicles (UAVs) provide flexible relay coverage when terrestrial infrastructure is unavailable or direct base station (BS)--user links are blocked. Low-altitude measurements at 2.4 and 5.9 GHz support a line-of-sight (LoS)-dominant Rician model around the 4-GHz carrier considered here \cite{ref1}. Amplify-and-forward (AF) relaying is attractive for low-complexity airborne platforms, but its end-to-end performance is jointly limited by the two hops and relay-noise amplification \cite{ref2}. With an array-equipped UAV, activating more elements increases receive and boresight gains but narrows the forwarding beam. When user locations are still uncertain, the resulting pointing loss can make full-array operation suboptimal.

Fluid antenna systems (FASs) \cite{wong1,wong2} provide a general reconfigurable-antenna paradigm in which the antenna configuration is treated as a controllable physical-layer resource rather than a fixed hardware constraint. This perspective expands the wireless design space by allowing the spatial response to adapt to propagation conditions, interference, and geometry. While early FAS research mainly focused on spatial diversity induced by small-scale fading \cite{fd0,fd1,fd1.5,fd2,fd2.5,fd3,fd3.5,fd4,fd5,fd5.5}, fluid antenna arrays (FAAs) \cite{FAA1,FAA2,FAA3} extend this reconfigurability to the array geometry itself. By controlling the effective aperture and spatial sampling structure, an FAA can directly shape beamwidth, directivity, sidelobes, and beamforming gain. 

The UAV relay considered in this work adopts an FAA-inspired architecture through the selection of a contiguous active subarray from a larger fixed antenna aperture. Although the physical array itself remains fixed, changing the number of active elements reconfigures the effective aperture and, consequently, its gain--beamwidth characteristics. This provides a practical form of geometric aperture adaptation for UAV relaying, in which the appropriate aperture depends not only on channel strength but also on the evolving reliability of user-location information. Unlike conventional fixed-array operation, the active subarray can expand as localization becomes more accurate and contract when directional uncertainty makes narrow-beam transmission unreliable.

Antenna activation in AF relays has mainly been optimized on the basis of instantaneous channel state information (CSI) \cite{CSI0,CSI1,CSI2,ref3} or long-term channel statistics \cite{ref4}, \cite{ref5}. UAV relay studies have also optimized trajectories or placements under known geometry \cite{ref6}, \cite{ref7}. In parallel, localization- and sensor-assisted beam control uses estimated positions, covariances, or angular uncertainty to adapt beamwidth and power \cite{ref8,ref9,ref10,ref11}. A jitter-aware UAV study \cite{ref12} further showed that transmitter-side attitude uncertainty can favor a finite antenna subset. These studies establish the value of uncertainty-aware adaptation, but they do not address a user-location posterior that contracts during flight and simultaneously drives the effective aperture of a subarray shared by both hops of an AF relay.

The key issue is how location uncertainty maps to array-direction uncertainty and then to reliable two-hop performance. A covariance matrix must be projected onto the array direction determined by the current UAV--user geometry \cite{ref13}, because equal position-error magnitudes can yield different pointing risks. From the FAA-inspired geometric perspective \cite{FAA1,FAA2,FAA3}, this projection determines how much effective aperture can be safely activated before the beam becomes excessively sensitive to location errors. The resulting gain--beamwidth tradeoff is especially relevant to the lower tail of the rate distribution, which motivates an empirical posterior $\varepsilon$-quantile rather than an average-rate criterion; this is consistent with outage-constrained communication design \cite{ref14}. Here, the UAV uses the posterior mean and covariance to characterize direction uncertainty, posterior samples to evaluate reliable rate, and long-term first-hop statistics to model the backhaul. These inputs are available at planning time; true positions and instantaneous fading are reserved for the execution audit.

This paper makes three contributions. First, covariance projection yields a closed-form direction-limited sizing law $M_{\mathrm{dir}}\propto1/\sigma_u$, explaining the transition from a finite effective aperture to full-array operation as the posterior contracts. Second, a minimum-power design jointly selects the common integer subarray size and UAV transmit power under a worst-user empirical $\varepsilon$-quantile rate requirement and total and per-element power limits. Third, a planning--execution audit evaluates the fixed posterior-based configuration over instantaneous two-hop Rician channels. At $t=8~\mathrm{s}$, the proposed design saves $3.17$ dB over full-array narrow-beam transmission and raises the service-success probability at a $0.15$-W budget from $43.2\%$ for a fixed eight-element subarray to $60.0\%$.

The remainder of this paper is organized as follows. Section II describes the two-hop AF relay geometry, the shared contiguous subarray, the channel model, and the location-posterior interface. Section III develops the FAA-inspired design, comprising the closed-form direction-limited sizing law, the posterior two-hop AF rate model, the minimum-power configuration search, and the instantaneous-channel execution audit. Section IV presents the simulation results, and Section V concludes the paper. Throughout, $(\cdot)^T$ and $(\cdot)^H$ denote transpose and conjugate transpose, $\|\cdot\|_2$ the Euclidean norm, $\mathbb{E}[\cdot]$ expectation, and $\mathcal{CN}(0,\sigma^2)$ the circularly symmetric complex Gaussian distribution. Boldface lowercase and uppercase letters denote vectors and matrices, respectively.

\section{System and Channel Model}
\label{sec:system}

\begin{figure}[!t]
\centering
\includegraphics[width=\columnwidth,keepaspectratio]{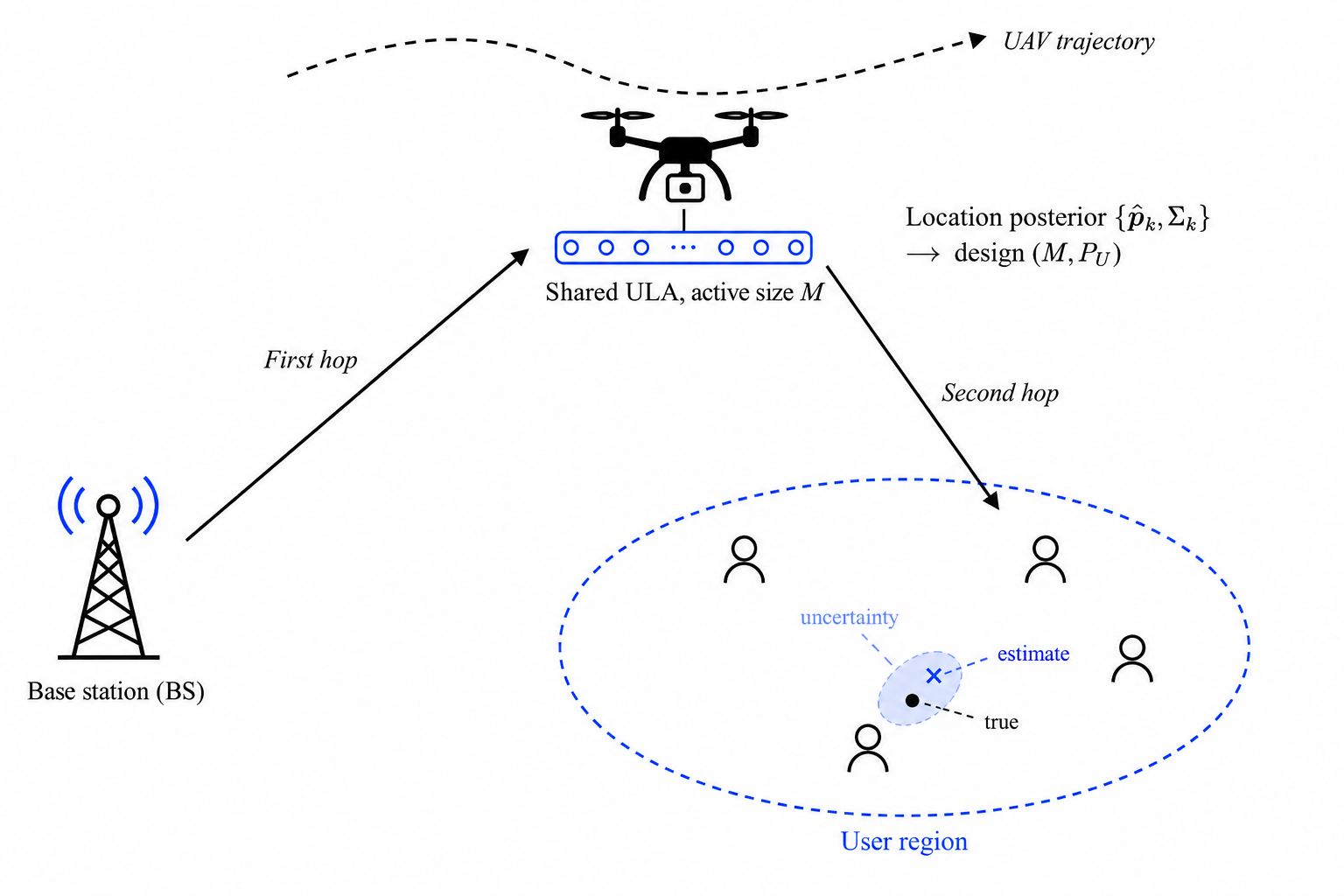}
\caption{Two-hop AF UAV relay with a shared active ULA and location-posterior-driven subarray and power design.}
\label{fig:1}
\end{figure}

Consider a two-hop AF relay with an $N_{\mathrm t}$-antenna BS, an $N_{\mathrm U}$-element UAV uniform linear array (ULA), and $K$ single-antenna users. Direct BS--user links are neglected. The UAV flies at fixed altitude $H$ along a prescribed trajectory $\mathbf q_n=[x_n,y_n,H]^T$; the BS is at $\mathbf q_{\mathrm B}$ and user $k$ at $\mathbf p_k=[x_k,y_k,0]^T$. Users are served on orthogonal resources and are coupled through a common active subarray size and the worst-user reliability requirement. Fig.~\ref{fig:1} summarizes the geometry and the posterior-aided configuration.

The users remain static during one mission. At time $n$, the localization interface provides

\begin{equation}
\mathcal B_n=\{\widehat{\mathbf p}_{k,n},\boldsymbol\Sigma_{k,n}\}_{k\in\mathcal K},
\label{eq:1}
\end{equation}

and a local Gaussian model $\mathbf p_k-\widehat{\mathbf p}_{k,n}\sim\mathcal N(\mathbf0,\boldsymbol\Sigma_{k,n})$ is used to draw $N_{\mathrm s}$ samples $\{\mathbf p_{k,n,j}\}$. The posterior mean and covariance are recursively estimated from matched-filtered uplink pilot observations using a particle filter (PF). The localization recursion itself lies outside the scope of the communication planner considered in the subsequent subarray-and-power design.

At each planning instant, a centered contiguous $M$-element subarray, $M\in\mathcal M\triangleq\{1,\ldots,N_{\mathrm U}\}$, is shared by first-hop reception and second-hop forwarding. The elements are half-wavelength spaced, and the ULA is fixed along the $y$-axis, perpendicular to the prescribed $x$-direction flight path. The relay is half-duplex, giving the pre-log factor $1/2$.

Both hops use narrowband Rician fading \cite{ref1}. With $\xi_{\mathrm{BU}}=c_{\mathrm{BU}}\|\mathbf q_n-\mathbf q_{\mathrm B}\|^{-\beta_{\mathrm{BU}}/2}$ and $\xi_k=c_k\|\mathbf q_n-\mathbf p_k\|^{-\beta_k/2}$,

\begin{equation}
\begin{aligned}
\mathbf H_{\mathrm{BU}}
&=\xi_{\mathrm{BU}}\Bigg(
\sqrt{\frac{\kappa_{\mathrm{BU}}}{\kappa_{\mathrm{BU}}+1}}
e^{j\theta_{\mathrm{BU}}}\mathbf A_{\mathrm{BU}}\\
&\hspace{18mm}+\sqrt{\frac{1}{\kappa_{\mathrm{BU}}+1}}
\mathbf X_{\mathrm{BU}}\Bigg),
\end{aligned}
\label{eq:2}
\end{equation}

\begin{equation}
\begin{aligned}
\mathbf h_k
&=\xi_k\Bigg(
\sqrt{\frac{\kappa_k}{\kappa_k+1}}e^{j\theta_k}
\mathbf a_{N_{\mathrm U}}(\vartheta_{k,n})\\
&\hspace{18mm}+\sqrt{\frac{1}{\kappa_k+1}}\mathbf g_k\Bigg),
\end{aligned}
\label{eq:3}
\end{equation}

where $\mathbf A_{\mathrm{BU}}=\mathbf a_{N_{\mathrm U}}(\vartheta_{\mathrm U,n})\mathbf a_{N_{\mathrm t}}^H(\vartheta_{\mathrm B,n})$, $\mathbf X_{\mathrm{BU}}$ and $\mathbf g_k$ have i.i.d. $\mathcal{CN}(0,1)$ entries, and the LoS phases are uniform on $[0,2\pi)$. For a half-wavelength-spaced ULA,

\begin{equation}
\mathbf a_N(\vartheta)=
[1,e^{-j\pi\sin\vartheta},\ldots,e^{-j\pi(N-1)\sin\vartheta}]^T.
\label{eq:4}
\end{equation}

Transmit and receive beams use the corresponding unit-norm steering vectors. Let $\mathbf f_{\mathrm B,n}=\mathbf a_{N_{\mathrm t}}(\vartheta_{\mathrm B,n})/\sqrt{N_{\mathrm t}}$ be the fixed LoS-matched BS beam, and let $\widetilde{\mathbf h}_{\mathrm{BU},n}$ denote the post-beamforming vector $\mathbf H_{\mathrm{BU}}\mathbf f_{\mathrm B,n}$ normalized to unit mean power per UAV element. Hence $\mathbb E[\|\mathbf S_M\widetilde{\mathbf h}_{\mathrm{BU},n}\|_2^2]=M$. The simulations generate this normalized vector directly as Rician with effective factor $\kappa_{\mathrm{BU},\mathrm{eff}}=10$; fixed BS power, beamforming gain, and UAV receiver noise are absorbed into the per-element first-hop SNR coefficient $\rho_{\mathrm{BU},n}$. Let $\mathbf S_M$ select the active contiguous subarray. Planning uses $\mathcal I_n=\{\mathbf q_n,\mathcal B_n,\mathcal S_{\mathrm{ch}}\}$, where $\mathcal S_{\mathrm{ch}}$ collects the long-term channel, noise, and power parameters available to the planner, and outputs $(\widehat M_n,\widehat P_{\mathrm U,n})$. True positions and instantaneous channels enter only the audit.

\section{Posterior-Driven Subarray and Power Design}
\label{sec:design}

\subsection{Direction Uncertainty and Subarray Gain}

For the $y$-axis ULA, define the user direction cosine

\begin{equation}
u_k(\mathbf q_n,\mathbf p_k)
=\frac{p_{k,y}-q_{n,y}}{\|\mathbf p_k-\mathbf q_n\|_2}
=\sin\vartheta_{k,n}.
\label{eq:5}
\end{equation}

The beam is steered to $\widehat u_{k,n}=u_k(\mathbf q_n,\widehat{\mathbf p}_{k,n})$. With $\widehat\vartheta_{k,n}=\arcsin(\widehat u_{k,n})$, the posterior-directed unit-norm forwarding beam is $\mathbf w_{k,n}(M)=\mathbf a_M(\widehat\vartheta_{k,n})/\sqrt M$. Linearizing around the posterior mean gives

\begin{equation}
\begin{aligned}
\Delta u_{k,n}
&\simeq\mathbf J_{k,n}(\mathbf p_k-\widehat{\mathbf p}_{k,n}),\\
\sigma_{u,k,n}^2
&\simeq\mathbf J_{k,n}\boldsymbol\Sigma_{k,n}\mathbf J_{k,n}^T,
\end{aligned}
\label{eq:6}
\end{equation}

where $\mathbf J_{k,n}=\nabla_{\mathbf p_k}u_k|_{\widehat{\mathbf p}_{k,n}}$. For an $M$-element unit-norm beam, the LoS power gain at direction offset $\Delta u$ is

\begin{equation}
G_M(\Delta u)=\frac{1}{M}\left|\sum_{m=0}^{M-1}e^{j\pi m\Delta u}\right|^2
=\frac{\sin^2(M\pi\Delta u/2)}{M\sin^2(\pi\Delta u/2)}.
\label{eq:7}
\end{equation}

Let $\delta_{k,n}=z_{1-\varepsilon/2}\sigma_{u,k,n}$. Because $\Delta u_{k,n}$ is Gaussian and $G_M$ decreases monotonically with $|\Delta u|$ over the main lobe, the $\varepsilon$-quantile gain is $G_M(\delta_{k,n})$. With $x=M\pi\delta_{k,n}/2$, maximizing this gain gives the stationarity condition $\tan x=2x$, whose first positive root $x_0\approx1.1656$ yields the direction-limited size

\begin{equation}
M_{k,n,\mathrm{dir}}
\simeq\frac{2x_0}{\pi\delta_{k,n}}
\approx\frac{0.742}{z_{1-\varepsilon/2}\sigma_{u,k,n}}.
\label{eq:dir-size}
\end{equation}

Since $M_{k,n,\mathrm{dir}}\delta_{k,n}\approx0.742<2$, the main-lobe approximation is self-consistent. The rule reflects the gain--beamwidth tradeoff directly: boresight gain grows with $M$, whereas the main-lobe width shrinks approximately as $1/M$, so the lower-tail gain has a finite maximizer. For $\varepsilon=0.05$, $z_{1-\varepsilon/2}=1.96$ and the rule reduces to $M_{k,n,\mathrm{dir}}\approx0.379/\sigma_{u,k,n}$. It is a direction-only reference: it evaluates the tail offset in angle but does not include the residual distance uncertainty that also perturbs second-hop path loss. The final integer $M$ is therefore determined by the complete two-hop reliable-rate design.

\subsection{Posterior Two-Hop AF Rate}

Overlined quantities denote planning-stage predictions; unmarked quantities in Section III-D denote instantaneous audit realizations. The first-hop statistical SNR is modeled as

\begin{equation}
\overline\gamma_{\mathrm{BU},n}(M)=\rho_{\mathrm{BU},n}M.
\label{eq:9}
\end{equation}

The per-element normalization gives $\mathbb E[\|\mathbf S_M\widetilde{\mathbf h}_{\mathrm{BU},n}\|_2^2]=M$. Thus, \eqref{eq:9} replaces the instantaneous combined first-hop gain by the mean available to the planner. Increasing $M$ strengthens this hop and the second-hop boresight gain, but also narrows $G_M(\cdot)$ and increases pointing sensitivity. The selected size therefore balances the two hops rather than optimizing either one in isolation.

For posterior sample $\mathbf p_{k,n,j}$, define $\Delta u_{k,n,j}=u_k(\mathbf q_n,\mathbf p_{k,n,j})-\widehat u_{k,n}$ and $\ell_{k,n,j}=c_k^2\|\mathbf q_n-\mathbf p_{k,n,j}\|^{-\beta_k}$. The predicted second-hop SNR is

\begin{equation}
\overline\gamma_{\mathrm U k,n,j}(M,P_{\mathrm U})
=\frac{P_{\mathrm U}\ell_{k,n,j}G_M(\Delta u_{k,n,j})}{\sigma_k^2}.
\label{eq:10}
\end{equation}

Using the standard two-hop AF relation \cite{ref2},

\begin{equation}
\begin{aligned}
\overline\gamma_{k,n,j}
&=\frac{\overline\gamma_{\mathrm{BU},n}\overline\gamma_{\mathrm U k,n,j}}
{\overline\gamma_{\mathrm{BU},n}+\overline\gamma_{\mathrm U k,n,j}+1},\\
\overline R_{k,n,j}
&=\frac12\log_2(1+\overline\gamma_{k,n,j}).
\end{aligned}
\label{eq:11}
\end{equation}

Planning uses posterior geometry and deterministic LoS array gain; small-scale fading is introduced only in the audit.

\subsection{Reliable Sizing and Power Design}

Let $Q_\varepsilon[\cdot]$ denote the empirical $\varepsilon$-quantile over the $N_{\mathrm s}$ posterior samples. Define

\begin{equation}
\begin{aligned}
\overline R_{k,n,\varepsilon}(M,P_{\mathrm U})
&=Q_\varepsilon\!\left[\{\overline R_{k,n,j}(M,P_{\mathrm U})\}_{j=1}^{N_{\mathrm s}}\right],\\
\overline J_n(M,P_{\mathrm U})
&=\min_{k\in\mathcal K}\overline R_{k,n,\varepsilon}(M,P_{\mathrm U}).
\end{aligned}
\label{eq:12}
\end{equation}

The configuration solves

\begin{equation}
\begin{gathered}
(\widehat M_n,\widehat P_{\mathrm U,n})\in\arg\min_{M,P_{\mathrm U}}\;P_{\mathrm U}\\
\mathrm{s.t.}\quad \overline J_n(M,P_{\mathrm U})\ge R_{\mathrm{req}},\quad M\in\mathcal M,\\
0\le P_{\mathrm U}\le P_{\mathrm U,\max},\quad P_{\mathrm U}/M\le P_{\mathrm{ant},\max}.
\end{gathered}
\label{eq:13}
\end{equation}

The quantile is taken after the two-hop AF rate composition, so it captures the joint effect of posterior distance and pointing uncertainty. The finite set $\mathcal M$ is searched directly. For each $M$, bisection finds the minimum feasible power because every sample rate, its empirical quantile, and hence $\overline J_n$ are nondecreasing in $P_{\mathrm U}$. The search stops when $10\log_{10}(P_{\mathrm{hi}}/P_{\mathrm{lo}})\le\tau_P$; numerically tied candidates favor the smaller $M$, which also provides a slightly wider beam. If no pair $(M,P_{\mathrm U})$ satisfies the constraints, the geometry is declared planning-infeasible; it is excluded from conditional power averages and counted as a service failure in Section IV. The complexity is $\mathcal O(|\mathcal M|KN_{\mathrm s}I_P)$, where $I_P$ is the number of bisection iterations.

\subsection{Instantaneous Two-Hop Audit}

After planning, $(\widehat M_n,\widehat P_{\mathrm U,n})$ and the user beams remain fixed. At execution, the UAV uses instantaneous first-hop combining and variable-gain forwarding, even though the plan was computed from $\rho_{\mathrm{BU},n}$ alone. For audit realization $c$,

\begin{equation}
\gamma_{\mathrm{BU},n,c}(M)
=\rho_{\mathrm{BU},n}\|\mathbf S_M\widetilde{\mathbf h}_{\mathrm{BU},n,c}\|_2^2,
\label{eq:14}
\end{equation}

\begin{equation}
\gamma_{\mathrm U k,n,c}(M,P_{\mathrm U})
=\frac{P_{\mathrm U}|\mathbf h_{k,n,c}^H\mathbf S_M^T\mathbf w_{k,n}(M)|^2}{\sigma_k^2}.
\label{eq:15}
\end{equation}

The resulting rate uses the same AF composition as above. Under an external power budget $P_{\mathrm U,\mathrm{bud}}$, service succeeds if $\widehat P_{\mathrm U,n}\le P_{\mathrm U,\mathrm{bud}}$ and $\min_kR_{k,n,c}\ge R_{\mathrm{req}}$.

\section{Simulation Results and Performance Analysis}
\label{sec:simulations}

\emph{Simulation setup.} The BS is at $(-300,0,30)$ m and the UAV follows $\mathbf q_n=[-300+10n,0,100]^T$ m, $n=1,\ldots,40$, so that time index $n$ corresponds to $t=n~\mathrm{s}$. The posterior is refreshed every $T_{\mathrm{upd}}=3~\mathrm{s}$ and the communication configuration is recomputed every second. Four users are drawn uniformly in a disk of radius $r_{\mathrm u}=100~\mathrm{m}$ centered at $(100,0,0)$ m with minimum separation $d_{\min}=10~\mathrm{m}$. Each sensing update uses two matched-filtered uplink channel-vector snapshots at $P_{\mathrm{pilot}}=0.05~\mathrm{W}$. The localization module uses $N_{\mathrm p}=500$ particles per user and supplies $N_{\mathrm s}=6000$ samples through the local Gaussian interface.

The first-hop per-element reference is $\gamma_{\mathrm{BU},0,\mathrm{el}}=20$ dB at $d_{\mathrm{BU},0}=200$ m, giving

\begin{equation}
\rho_{\mathrm{BU},n}=\gamma_{\mathrm{BU},0,\mathrm{el}}
\left(\frac{d_{\mathrm{BU},0}}{\|\mathbf q_n-\mathbf q_{\mathrm B}\|_2}\right)^{\beta_{\mathrm{BU}}}.
\label{eq:16}
\end{equation}

The second-hop reference is $\gamma_{\mathrm U k,0}=0$ dB at $d_{\mathrm U k,0}=150$ m for $P_{\mathrm{ref}}=0.1$ W and unit array gain. Table~\ref{tab:parameters} lists the remaining parameters.

\begin{table}[!t]
\caption{Main Simulation Parameters}
\label{tab:parameters}
\centering
\footnotesize
\setlength{\tabcolsep}{2.2pt}
\renewcommand{\arraystretch}{1.08}
\begin{tabularx}{\columnwidth}{@{}>{\raggedright\arraybackslash}X >{\raggedright\arraybackslash}p{0.19\columnwidth} >{\raggedright\arraybackslash}X >{\raggedright\arraybackslash}p{0.19\columnwidth}@{}}
\toprule
\textbf{Parameter} & \textbf{Value} & \textbf{Parameter} & \textbf{Value} \\
\midrule
$K$ & 4 & $N_{\mathrm U}$ & 16 \\
$f_{\mathrm c}$ & 4 GHz & $H$ & 100 m \\
$r_{\mathrm u}$ & 100 m & $d_{\min}$ & 10 m \\
$\beta_{\mathrm{BU}}$ & 2.0 & $\beta_k$ & 2.2 \\
$\varepsilon$ & 0.05 & $R_{\mathrm{req}}$ & 0.4 bit/s/Hz \\
$P_{\mathrm U,\max}$ & 1 W & $P_{\mathrm{ant},\max}$ & 0.0625 W \\
$N_{\mathrm s}$ & 6000 & $N_{\mathrm p}$ & 500 per user \\
$\kappa_{\mathrm{BU},\mathrm{eff}}$ & 10 & $\kappa_k$ & 10 \\
$\tau_P$ & 0.01 dB & $T_{\mathrm{upd}}$ & 3 s \\
$P_{\mathrm{pilot}}$ & 0.05 W & $N_{\mathrm{dyn}}$ & 50 \\
$N_{\mathrm{plan}}$ & 200 & $N_{\mathrm C}$ & 10 \\
\bottomrule
\end{tabularx}
\end{table}

The compared schemes are Proposed, Full narrow ($M=16$), Full wide (optimized), and Fixed subarray $M=8$. Full wide (optimized) keeps all $16$ elements active and selects from $\mathcal D_{\mathrm{wide}}=\{1,3,5,7,9,11,13\}$, where each value denotes the number of synthesis directions used by the phase-synthesis broadening routine and $1$ recovers the narrow beam. The baseline therefore tests beam broadening without reducing the active aperture. Dynamic curves average $N_{\mathrm{dyn}}=50$ geometries; power and service audits use $N_{\mathrm{plan}}=200$ geometries and $N_{\mathrm C}=10$ paired instantaneous channel realizations per geometry. The position RMSE decreases from $26.6$ m at $5$ s to $16.3$ m at $15~\mathrm{s}$, while the mean directional standard deviation decreases from $0.0893$ to $0.0240$. The stronger contraction of $\sigma_u$ reflects not only posterior shrinkage but also the time-varying Jacobian that projects the position covariance onto the array direction; position RMSE alone therefore does not determine pointing risk.

\begin{figure}[!t]
\centering
\includegraphics[width=\columnwidth,keepaspectratio]{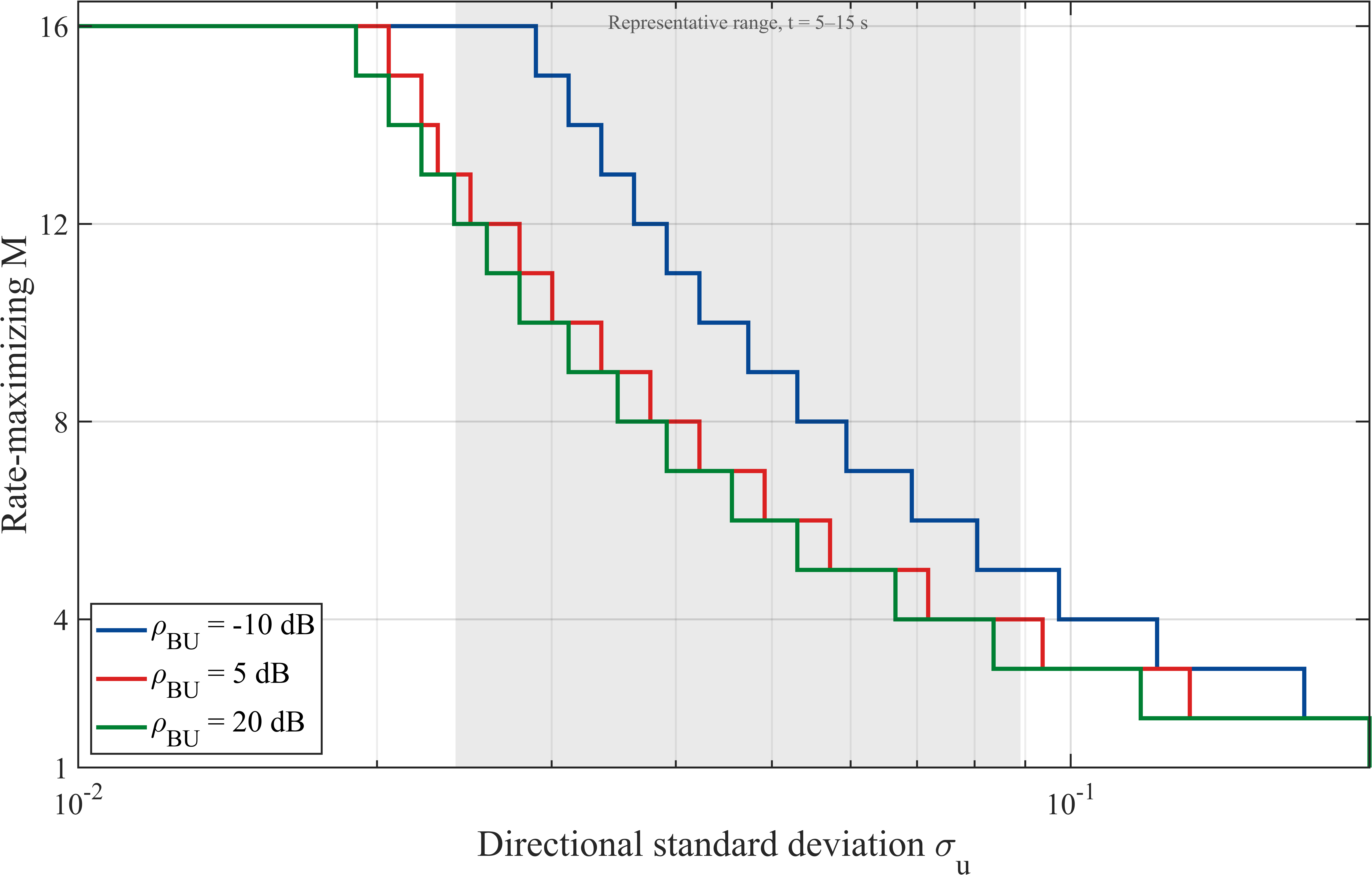}
\caption{Rate-maximizing $M$ versus directional standard deviation in the auxiliary fixed-power experiment, with $\rho_{\mathrm{BU}}\in\{-10,5,20\}$ dB. The shaded band marks the $\sigma_u$ range observed from $t=5$ to $13~\mathrm{s}$ during the corresponding dynamic adaptation experiment shown in Fig.~\ref{fig:3}.}
\label{fig:2}
\end{figure}

\emph{Selection mechanism.} Fig.~\ref{fig:2} isolates the sizing mechanism by maximizing rate at fixed $P_{\mathrm U}=0.1$ W with power caps disabled. The four users have nominal second-hop SNRs $\{2,1,0,-1\}$ dB, and the fixed per-element first-hop coefficient is swept over $\rho_{\mathrm{BU}}\in\{-10,5,20\}$ dB. Larger $\sigma_u$ favors smaller $M$ because a wider beam reduces pointing loss; under weaker first-hop conditions, more elements are retained because the receive-array gain in \eqref{eq:9} remains valuable. The $5$- and $20$-dB curves nearly coincide once the first hop ceases to limit the AF composition, which is consistent with the two-hop balance in Section III-B.

\suppressfloats[t]
\begin{figure}[t]
\centering
\includegraphics[width=\columnwidth,keepaspectratio]{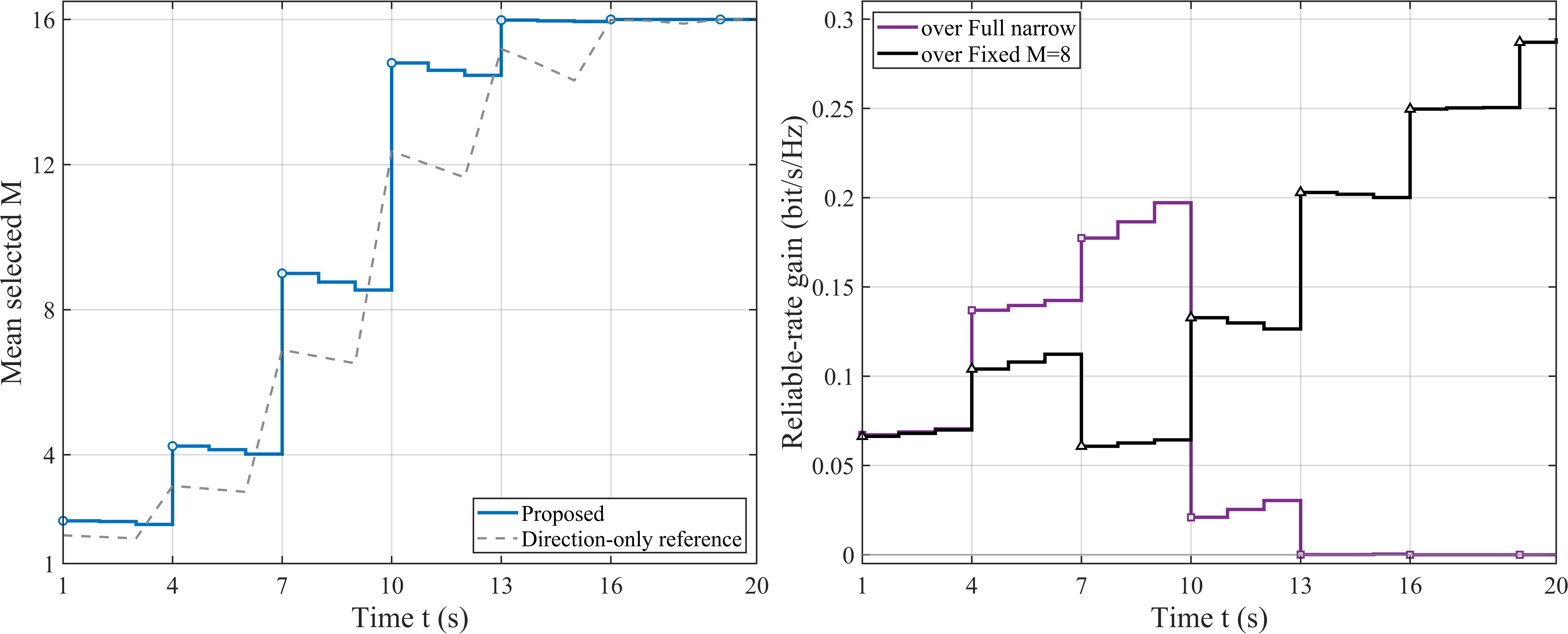}
\par\vspace{0.2em}
\begin{minipage}[t]{0.48\columnwidth}
\centering
\footnotesize (a)
\end{minipage}
\hfill
\begin{minipage}[t]{0.48\columnwidth}
\centering
\footnotesize (b)
\end{minipage}
\caption{Dynamic adaptation at each planning instant: (a) mean selected subarray size and the direction-only reference; (b) posterior $5\%$-quantile rate gain of Proposed over the Full narrow and Fixed subarray $M=8$ baselines.}
\label{fig:3}
\end{figure}

\begin{figure}[!t]
\centering
\vspace{-\floatsep}
\vspace{2pt}
\includegraphics[width=\columnwidth,keepaspectratio]{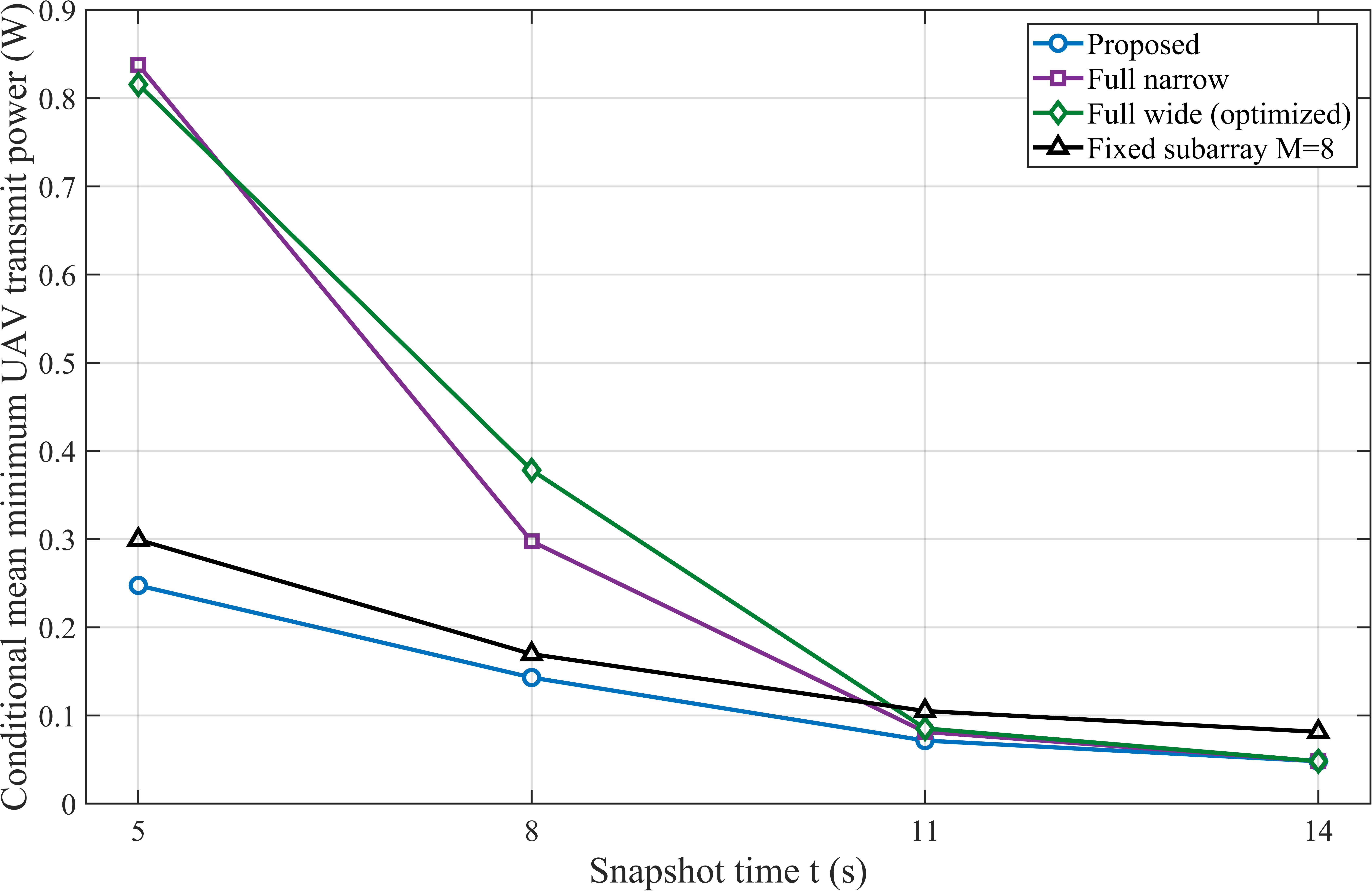}
\caption{Conditional mean minimum UAV power at $t\in\{5,8,11,14\}~\mathrm{s}$.}
\label{fig:4}
\end{figure}

\begin{figure}[!t]
\centering
\includegraphics[width=\columnwidth,keepaspectratio]{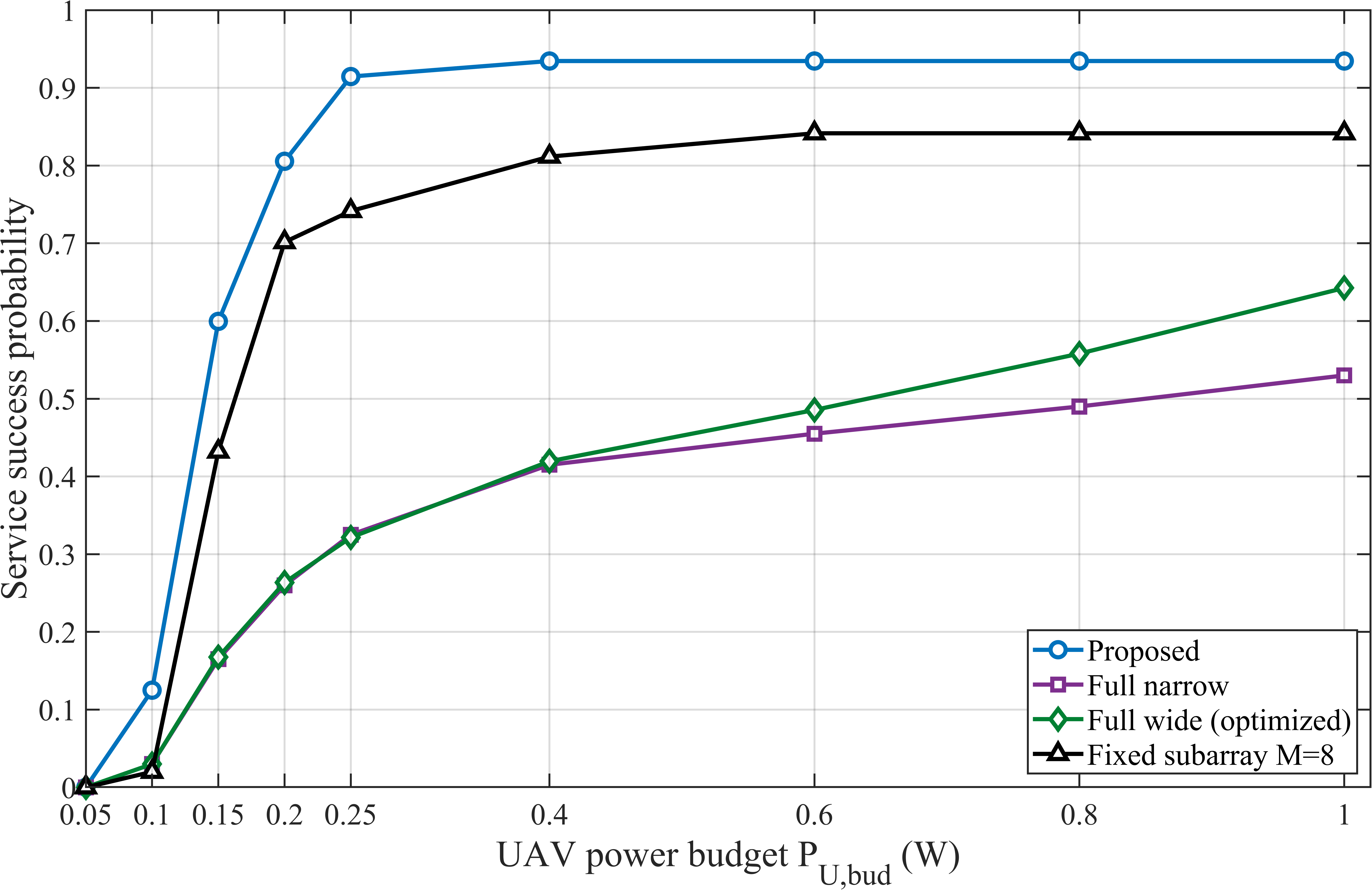}
\caption{Service-success probability versus UAV power budget at the $t=8~\mathrm{s}$ snapshot under posterior-planned configurations and instantaneous two-hop Rician channels.}
\label{fig:5}
\end{figure}

\emph{Dynamic adaptation.} Fig.~\ref{fig:3}(a) shows upward transitions after the $3$-s posterior updates. Applying the closed-form rule to the limiting user gives the continuous reference

\begin{equation}
M_{n,g,\mathrm{dir}}=\min\left\{N_{\mathrm U},\frac{0.379}{\max_k\sigma_{u,k,n,g}}\right\}.
\label{eq:direction-reference}
\end{equation}

This direction-only reference is plotted without integer rounding and is used only as a continuous benchmark. It captures the trend, while the complete rate-quantile design selects slightly larger subarrays because distance and direction errors are jointly propagated. At $t=8~\mathrm{s}$, the mean sizes are $8.76$ for Proposed and $6.71$ for the direction-only reference; both approach $M=16$ after $13~\mathrm{s}$. The rule predicts a full-array directional threshold of $0.379/N_{\mathrm U}=0.0237$, close to the observed mean $\sigma_u=0.0240$ at $15~\mathrm{s}$, which is consistent with the transition scale. Fig.~\ref{fig:3}(b) shows reliable-rate gains of $0.187$ and $0.063$ bit/s/Hz over Full narrow and Fixed subarray $M=8$ at $t=8~\mathrm{s}$. The gain over Full narrow vanishes after posterior convergence, whereas the fixed subarray becomes increasingly gain-limited. Doubling $P_{\mathrm{ant},\max}$ changes mean selected $M$ by at most $0.14$, confirming that the observed adaptation is not driven by the per-element cap. Only the first $20~\mathrm{s}$ are shown because the selected configuration remains at the full array thereafter throughout the remainder of the simulated trajectory.

\emph{Minimum required power.} Fig.~\ref{fig:4} shows the conditional mean minimum UAV power. At $t=8~\mathrm{s}$, Proposed requires $0.143$ W, versus $0.298$, $0.378$, and $0.170$ W for Full narrow, Full wide (optimized), and Fixed subarray $M=8$, with feasibility probabilities $0.940$, $0.530$, $0.675$, and $0.845$, respectively. Full wide (optimized) covers more difficult geometries than Full narrow, so its conditional mean is evaluated over a broader feasible set and can be higher despite the added broadening options. On geometries feasible for both Proposed and Full narrow, the paired saving is $0.184$ W ($3.17$ dB), with a $95\%$ geometry-cluster-bootstrap interval $[0.143,0.229]$ W. By $t=14~\mathrm{s}$, Proposed and the full-array schemes require about $0.048$ W, while Fixed subarray $M=8$ requires $0.082$ W.

\emph{Instantaneous-CSI service audit.} Fig.~\ref{fig:5} evaluates the fixed $t=8~\mathrm{s}$ plans, counting planning-infeasible geometries as failures. At a $0.15$-W budget, service-success probabilities are $60.0\%$, $43.2\%$, $16.5\%$, and $14.7\%$ for Proposed, Fixed subarray $M=8$, Full narrow, and Full wide (optimized); at $0.25$ W, Proposed and Fixed subarray $M=8$ reach $91.5\%$ and $74.2\%$. Proposed saturates at $93.5\%$, close to its $94.0\%$ planning feasibility. Conditional audit outage is $0.6\%$ for Proposed, $4.7\%$ for Full wide (optimized), and $0.4\%$ for Fixed subarray $M=8$, with no outage observed for Full narrow; all are within the nominal $\varepsilon=5\%$ tail level. The conditional zero for Full narrow should be interpreted together with its much lower planning feasibility, rather than as evidence of higher overall service reliability across the full set of tested geometries.

\section{Conclusion}
\label{sec:conclusion}
This paper developed an FAA-inspired, location-posterior-driven subarray sizing and power-control framework for a two-hop AF UAV relay. By adaptively selecting the active contiguous subarray, the UAV reconfigures its effective aperture to balance receive and forwarding gains against beamwidth robustness. Covariance projection gives the closed-form trend $M_{\mathrm{dir}}\propto1/\sigma_u$, while posterior rate samples determine the minimum-power configuration satisfying the reliable-rate requirement. Geometric aperture adaptation is most beneficial during the intermediate-uncertainty regime and gradually converges to full-array operation as the location posterior contracts. At $t=8~\mathrm{s}$, the proposed design saves $3.17$ dB over full-array narrow-beam transmission and raises the $0.15$-W service-success probability from $43.2\%$ for a fixed eight-element subarray to $60.0\%$. Future work will consider joint trajectory and aperture design, mobile users, non-Gaussian posteriors, and physically reconfigurable FAA implementations.

\end{document}